\documentstyle[12pt]{article}
\date{}
\def\be{\begin{equation}}
\def\ee{\end{equation}}
\def\bea{\begin{eqnarray}}
\def\eea{\end{eqnarray}}
\def\s{\sigma}
\def\al{\alpha}

\def\de{\delta}
\def\om{\omega}
\def\pr{\prime}
\def\th{\theta}
\def\fpar{f_\parallel}

\title{ Quasirotational disturbances\\
of the relativistic string with massive ends\\
and higher radial excitations of hadrons}
\author{G.\,S. Sharov\\%\thanks{E-mail: german.sharov@tversu.ru}\\
{\small Tver state university}\\
{\small Tver, 170002, Sadovyj per. 35, Mathem. dep-t.}}
\begin{document}
\maketitle
\begin{abstract}
For the relativistic string with massive ends the small disturbances
of its rotational motion (quasirotational states) are investigated.
They are presented in the form of Fourier series with the two
types of oscillatory modes. They have the form of stationary
waves correspondently in the rotational plane and in the orthogonal direction.
The calculated values of the energy and angular momentum
of these states give us possibility to describe higher radial excitations
for hadrons with the help of the planar quasirotational oscillatory modes
of the string with massive ends.
\end{abstract}

\bigskip
\noindent{\bf Introduction}
\medskip

The relativistic string with massive ends \cite{Ch,BN} includes two massive
points connected by the string that simulates the QCD confinement
mechanism. This system may be considered as the string model of the meson
$q$-$\overline q$ or the quark-diquark model $q$-$qq$ of the baryon \cite{Ko}.

This model in the natural way describes the (quasi)linear Regge
trajectories for the orbital excitations of mesons and baryons. For this
purpose different authors use the rotational motions (planar uniform rotations
of the rectilinear string) \cite{Ko,Solov,4B,InSh}. But for describing
the higher radial excitations and other hadron excited states we are to use
more complicated string motions than rotational ones. This problem of
widening the application area of the string model onto various types of
hadron excited states has not been solved yet in the framework of the
relativistic string with massive ends.

From this point of view the small disturbances of rotational motions
(quasirotational states) of this system obtained in Ref.~\cite{stabPRD}
are very interesting. In this paper
after the brief review the classical dynamics for the relativistic string
with massive ends in Sect.~1 we consider in Sect.~2 these quasirotational
states. In Sect.~3 possibilities of application these states
for describing the higher radial excitations of hadrons are investigated.

\bigskip
\noindent{\bf 1. Dynamics and rotational motions of the string
with massive ends}
\medskip

The relativistic string with the pointlike masses $m_1$, $m_2$
at the ends of the string is described by the action \cite{Ch}
\be
 S=-\int\limits_{\tau_1}^{\tau_2}\! d\tau\left\{\gamma\!
\int\limits_{\s_1(\tau)}^{\s_2(\tau)}\!\!
\left[\big(\dot X,X'\big)^2-\dot X^2X'{}^2\right]^{1/2}\!d\s+\sum
_{i=1}^3m_i\sqrt{\dot x_i^2(\tau)}\right\}.
\label{S}\ee
Here $X^\mu(\tau,\s)$ are coordinates of a point of
the string in $3+1$\,-\,dimensional\footnote{This approach is applicable
for an arbitrary dimensionality of $R^{1,D-1}$.} Minkowski space $R^{1,3}$
with signature $(+,-,-,\dots)$ and the (pseudo)scalar product
$\big( a,b\big)=a^\mu b_\mu$, $\gamma$ is the string tension,
the speed of light $c=1$,
$\dot X^\mu=\partial_\tau X^\mu$, $X'{}^\mu=\partial_\s X^\mu$,
$\dot x_i^\mu(\tau)=\frac d{d\tau}X^\mu(\tau,\s_i(\tau))$; $\s_i(\tau)$
are inner coordinates of endpoints' world lines. These massive points
simulate the quark-antiquark pair for the meson or the quark and diquark
for the baryon model $q$-$qq$.

The equations of motion and the boundary conditions at the ends
result from the action (\ref{S}) and take the following simplest form
\bea
&\ddot X^\mu-X''{}^\mu=0,&\label{eq}\\
&\displaystyle m_i\frac d{d\tau}U^\mu_i(\tau)+(-1)^i\gamma
X'{}^\mu(\tau,\s_i)=0,\quad\;\; U^\mu_i(\tau)=\frac{\dot x_i^\mu(\tau)}
{\dot x_i^\mu(\tau)}=\frac{\dot X^\mu(\tau,\s_i)}{|\dot X(\tau,\s_i)|},&\!\!
\label{qq}\eea
under the conditions $\s_1=0$, $\s_2=\pi$ for the ends
and the orthonormality conditions on the world surface
\be
\dot X^2+X'{}^2=0,\qquad(\dot X,X')=0,
\label{ort}\ee
which always may be stated without loss of generality \cite{Ch,stabPRD}.
Under conditions (\ref{ort}) the equations of motion (\ref{eq}) become linear
but the boundary conditions (\ref{qq}) for the massive points
remain essentially nonlinear.
They make the model much more realistic
but they bring additional nonlinearity and (hence) a lot of problems
with quantization of this model.

For the relativistic string with massive ends the classical solution
describing the rotational motion (planar uniform rotation of the
rectilinear string segment) is well known \cite{Ch,Ko} and may be represented
in the form \cite{4B,stabPRD}
\be
X^\mu(\tau,\s)=X^\mu_{rot}(\tau,\s)=\Omega^{-1}\big[\th\tau e_0^\mu+
\cos(\th\s+\phi_1)\cdot e^\mu(\tau)\big].
\label{rot}\ee
Here $\Omega$ is the angular velocity,
$e_0^\mu$ is the unit time-like velocity vector of c.m.,
\be
e^\mu(\tau)=e_1^\mu\cos\th\tau+e_2^\mu\sin\th\tau
\label{evec}\ee
is the unit ($e^2=-1$)
space-like rotating vector directed along the string,
$\s\in[0,\pi]$. The parameter $\th$ (dimensionless frequency) is connected
with the constant speeds $v_1$, $v_2$ of the ends through the relations
\be
v_1=\cos\phi_1,\quad v_2=-\cos(\pi\th+\phi_1),\quad
\frac{m_i\Omega}\gamma=\frac{1-v_i^2}{v_i}.
\label{v12Om}\ee

Expression (\ref{rot}) under restrictions (\ref{v12Om}) is the exact solution of the classic
equations of motion (\ref{eq}) and satisfies the orthonormality (\ref{ort}) and boundary conditions (\ref{qq}).
The classic expressions for the energy $E$ and the angular momentum $J$ (its projection
onto $Oz$ or $e_3^\mu$-direction) of the states (\ref{rot}) are \cite{Ch,Ko,4B}
\be
E_{rot}=\sum_{i=1}^2\bigg[\frac{\gamma\arcsin v_i}\Omega+\frac{m_i}\
{\sqrt{1-v_i^2}}\bigg],\quad
J_{rot}=\frac1{2\Omega}\bigg\{\sum_{i=1}^2\bigg[\frac{\gamma\arcsin v_i}\Omega+
\frac{m_iv_i^2}{\sqrt{1-v_i^2}}\bigg]\bigg\},
\label{EJ}\ee
The implicit expression (\ref{EJ}) with different form of taking into account quark spins
and the spin-orbit correction \cite{Ko,4B} describes quasilinear Regge trajectories
$J=J(E^2)$ with the ultrarelativistic behavior
$J\simeq\alpha'E^2-\alpha_1 E^{1/2}$, $E\to\infty$,
where the slope has the Nambu value $\alpha'=(2\pi\gamma)^{-1}$.
So the rotational motions  of the string models $q$-$\overline q$ and $q$-$qq$
are widely used for modeling the orbitally excited hadron
states on the leading (parent) Regge trajectories \cite{Ko,Solov,4B,InSh}.

But other excited hadron states, for example, the radial excitations on the
daughter Regge trajectories \cite{InSh} are so far beyond the field
of application for the string model (\ref{S}). There were many attempts
to specify the motions of the relativistic string (\ref{S}), which may be
interpreted as the radial excitations of hadrons. In particular,
for this purpose small disturbances of the rotational motion (\ref{rot})
were searched in Refs.~\cite{Ida,AllenOV}. But these attempts were not
fruitful. It was shown in Ref.~\cite{Nestrad} that these disturbances do not
satisfy the rectilinearity ansatz \cite{Ida} in the string dynamics.
The approach in Refs.~\cite{AllenOV} involves the complicated nonlinear
form of string motion equations beyond the conditions (\ref{ort})
and includes some oversimplified assumptions, in particular, neglecting
the boundary condition (\ref{qq}) at the moving end (the string with one fixed
end was considered).
So the solutions \cite{AllenOV} were not correct (details are in Ref.~\cite{stabPRD}).

In Ref.~\cite{stabPRD} another approach for obtaining small disturbances
of the rotational motion\footnote{We use below the term ``quasirotational
states" for these disturbances.} (\ref{rot}) was suggested for
the relativistic string with massive and fixed ends ($0<m_1<\infty$,
$m_2\to\infty$). In the next section this approach is generalized
for the case of two finite masses $m_1$, $m_2$ at the ends of the string.

\bigskip
\noindent{\bf 2. Quasirotational states for the meson string model}
\medskip

The quasirotational states or small disturbances of the rotational motion
(\ref{rot}) are interesting due to the following three reasons:
(a) we are to search the motions describing the radially excited hadron states,
in other words, we are to describe the daughter Regge trajectories;
(b) the quasirotational motions are necessary for solving the problem
of stability of rotational states (\ref{rot});

(c) the quasirotational states are the basis for quantization of these
nonlinear problems in the linear vicinity of the solutions (\ref{rot})
(if they are stable).

For obtaining the quasirotational solutions of slightly curved
string (\ref{S}) the following approach was suggested in Ref.~\cite{stabPRD}.
Under the orthonormality conditions (\ref{ort}) we use
the linear equations of motion (\ref{eq}) and substitute
their general solution
\be
X^\mu(\tau,\s)=\frac1{2}\Big[\Psi^\mu_+(\tau+\s)+\Psi^\mu_-(\tau-\s)\Big]
\label{gen}\ee
into the boundary conditions (\ref{qq}).
So the problem is reduced to the system of ordinary differential equations
with shifted arguments.
The unknown function may be $\Psi^\mu_+(\tau)$, $\Psi^\mu_-(\tau)$, or
unit velocity vectors (\ref{qq}) of the endpoints $U^\mu_1(\tau)$ or $U^\mu_2(\tau)$
--- this is equivalent due to the relations \cite{PeSh}
\be
\Psi^{\pr\mu}_\pm(\tau\pm\s_i)=m_i\gamma^{-1}\Big[
\sqrt{-U_i^{\pr2}(\tau)}\,U_i^\mu(\tau)\mp(-1)^i U_i^{\pr\mu}(\tau)\Big].
\label{psdet}\ee
Remind that $\s_1=0$, $\s_2=\pi$.

Taking relations (\ref{ort}) and (\ref{gen}) into account we transform
the boundary conditions (\ref{qq}) into the systems \cite{PeSh}
$$
\frac{dU_i^\mu}{d\tau}=\mp(-1)^i\frac\gamma{m_1}\big[\de^\mu_\nu-
U_i^\mu(\tau)\,U_{i\nu}(\tau)\big]\,\Psi^{\pr\nu}_\pm(\tau\pm\s_i),
$$
where $\de^\mu_\nu=\left\{\begin{array}{cl} 1, & \mu=\nu\\ 0, & \mu\ne\nu.
\end{array}\right.$

Substituting Eqs.~(\ref{psdet}) into these relations we obtain for the case
$0<m_i<\infty$ the system
\be
\begin{array}{c}
U^{\pr\mu}_1(\tau)=m_2m_1^{-1}\big[\de^\mu_\nu-
U_1^\mu(\tau)\,U_{1\nu}(\tau)\big]\big[\sqrt{-U_2^{\pr2}(\tau-\pi)}\,
U_2^\nu(\tau-\pi)-U_2^{\pr\nu}(\tau-\pi)\big],\\
U^{\pr\mu}_2(\tau)=m_1m_2^{-1}\big[\de^\mu_\nu-
U_2^\mu(\tau)\,U_{2\nu}(\tau)\big] \big[\sqrt{-U_1^{\pr2}(\tau-\pi)}\,
U_1^\nu(\tau-\pi)-U_1^{\pr\nu}(\tau-\pi)\big],
\rule[3.5mm]{0mm}{1mm}\end{array}
\label{sysU}\ee
that plays the role of the above mentioned system of ordinary differential equations
with shifted arguments with respect to unknown  vector functions $U^\mu_i(\tau)$.
Note that the initial data (keeping all information about the string's motion)
may be given here as the function $U^\mu_1(\tau)$ or $U^\mu_2(\tau)$
in the segment $I=[\tau_0,\tau_0+2\pi]$ with the parameters $m_i/\gamma$,
$U^\mu_i(\tau_0+\pi)$. Integration of the system (\ref{sysU}) with this initial
data yields the values $U^\mu_1(\tau)$ and $U^\mu_2(\tau)$ for all $\tau$
and then we can obtain the world surface $ X^\mu(\tau,\s)$ with the help of
Eqs.~(\ref{psdet}) and (\ref{gen}).

For the rotational motion (\ref{rot}) the unit velocity vectors $U_i^\mu$ are
\be
U_i^\mu(\tau)=U_{i(rot)}^\mu(\tau)=\Gamma_i\big[e_0^\mu-(-1)^iv_i
\acute e^\mu(\tau)\big],\qquad\quad\Gamma_i=(1-v_i^2)^{-1/2},
\label{Urot}\ee
where $\acute e^\mu(\tau)=-e_1^\mu\sin\th\tau+e_2^\mu\cos\th\tau=
\th^{-1}\frac d{d\tau}e^\mu(\tau)$ is the unit rotating vector, connected
with the vector $e^\mu$ (\ref{evec}). Expressions (\ref{Urot}) are solutions
of the system (\ref{sysU}) if the parameters $v_i$, $m_i$, $\theta$ are
related by Eqs.~(\ref{v12Om}).

To study the small disturbances of the rotational motion (\ref{rot})
we consider arbitrary small disturbances of this motion or of the vectors
(\ref{Urot}) in the form
\be
U^\mu_i(\tau)=U^\mu_{i(rot)}(\tau)+u_i^\mu(\tau),\qquad |u_i^\mu|\ll1.
\label{U+u}\ee
For the exhaustive description of this quasirotational state the disturbance
$u_i^\mu(\tau)$ may be given in the initial segment $I=[\tau_0,\tau_0+2\pi]$.
It is small so we neglect in the linear approximation the second order terms.
The equality $U_i^2(\tau)=1$ for both vectors $U_i^\mu$ and $U_{i(rot)}^\mu$
leads in the linear approximation to the condition
\be
U_{i(rot)}^\mu(\tau)\,u_{i\mu}(\tau)=0.
\label{Uu}\ee

When we substitute the expressions (\ref{U+u}) into the system (\ref{sysU})
and omit the second order terms we obtain the linearized system of equations
describing the evolution of small arbitrary disturbances $u_i^\mu$.
Considering projections of these two vector equations onto the basic vectors
$e_0$, $e$, $\acute e$, $e_3$, we reduce it to the following system of
equations with respect to projections of $u_i^\mu$:
\be
\!\!\!\!\!\!\begin{array}{c}
u'_{10}(\tau)+Q_1u_{10}(\tau)-\Gamma_1Q_1u_{1e}(\tau)=
M_0\big[u'_{20}-Q_2u_{20}+\Gamma_2Q_2u_{2e}\big],\\
\!\!u'_{1e}(\tau)+Q_1u_{1e}(\tau)+\theta v_1^{-1}u_{10}(\tau)=
M_1^{-1}\big[-u'_{2e}-Q_1u_{2e}+N_2^*u'_{20}+N_2u_{20}\big],
\rule[3.3mm]{0mm}{1mm}\!\!\\
u'_{20}+Q_2u_{20}+\Gamma_2Q_2u_{2e}=
M_0^{-1}\big[u'_{10}(-)-Q_1u_{10}(-)-\Gamma_1Q_1u_{1e}(-)\big],
\rule[3.3mm]{0mm}{1mm}\\
\!\!\!\!u'_{2e}+Q_2u_{2e}-\theta v_2^{-1}u_{20}=
M_1\big[-u'_{1e}(-)-Q_2u_{1e}(-)+N_1^*u'_{10}(-)+N_1u_{10}(-)\big],
\rule[3.3mm]{0mm}{1mm}\!\!\\
u'_{1z}(\tau)+Q_1u_{1z}(\tau)=(m_2/m_1)\big[-u'_{2z}+Q_2u_{2z}\big],
\rule[3.3mm]{0mm}{1mm}\\
u'_{2z}+Q_2u_{2z}=(m_1/m_2)\big[-u'_{1z}(-)+Q_1u_{1z}(-)\big].
\rule[3.3mm]{0mm}{1mm}\end{array}\!\!\!
\label{sysu}\ee
Here $Q_i=\Gamma_i\theta v_i={}$const, $(-)\equiv(\tau-2\pi)$, the functions
\be
u_{i0}(\tau)=(e_0,u_i)=e_0^\mu u_{i\mu},\qquad u_{ie}(\tau)=(e,u_i),\qquad
u_{iz}(\tau)=(e_3,u_i)
\label{u0ez}\ee
are the projections of the vectors $u_i^\mu(\tau)$ onto the mentioned basis.
The projections of $u_i^\mu$ onto $\acute e^\mu$ may be expressed through
$u_{i0}$: $(\acute e,u_i)=(-1)^iv_i^{-1}u_{i0}$ due to the equality
(\ref{Uu}). Arguments $(\tau-\pi)$ of the functions $u_{20}$, $u_{2e}$,
$u_{2z}$ are omitted.
The constants in Eqs.~(\ref{sysu}) are
$$\begin{array}{c}
M_0=m_2Q_1/(m_1Q_2),\qquad M_1=m_1\Gamma_1/(m_2\Gamma_2),\\
N_i^*=-(-1)^i(1+Q_{3-i}/Q_i)/\Gamma_i,\qquad
N_i=(-1)^i(Q_{3-i}+Q_i\kappa_i)/\Gamma_i,\qquad\kappa_i=1+v_i^{-2}.
\rule[3.3mm]{0mm}{1mm}\end{array}$$

We shall search solutions of the linearized system (\ref{sysu}) in the form
\be
u_i^\mu=A_i^\mu e^{-i\om\tau}.
\label{uharm}\ee
For the last two equations (\ref{sysu}) (they form the closed subsystem)
solutions in the form (\ref{uharm}) exist only if the dimensionless
frequency $\om$ satisfies the transcendental equation
\be
\frac{\om^2-Q_1Q_2}{(Q_1+Q_2)\,\omega}=\cot\pi\omega.
\label{zfreq}\ee
Equation (\ref{zfreq}) has the countable set of real roots
$\om_n$, $n-1<\om_n<n$, the minimal positive root $\om_1$ is equal to the parameter
$\theta$ in Eq.~(\ref{rot}).
These pure harmonic $z$-disturbances corresponding to various $\om_n$
result in the following correction to the motion (\ref{rot}) [due to
Eqs.~(\ref{psdet}), (\ref{gen}) there is only $z$ or $e_3^\mu$\
component of the correction]:
\be
X^\mu(\tau,\s)=X_{rot}^\mu(\tau,\s)+e_3^\mu
A\cos(\om_n\s+\phi_n)\cdot\cos(\om_n\tau+\varphi_0),
\label{zwave}\ee
Here the amplitude $A$ is to be small in comparison with $\Omega^{-1}$.

Expression (\ref{zwave}) describes oscillating string in the form of
orthogonal (with respect to the rotational plane) stationary waves with
$n$ nodes in the interval $0<\s<\pi$. Note that the string ends are not in
nodes, they oscillate along $z$-axis at the frequency
$\Omega_n=\Omega\om_n/\theta$. The shape
$F=A\cos(\om_n\s+\phi_n)$ of the
$z$-oscillation (\ref{zwave}) is not pure sinusoidal with
respect to the distance $s=\Omega^{-1}\cos(\theta\s+\phi_1)$
from the center to a point ``$\s$":
If $n=1$ this dependence is linear. In this trivial case the motion is
pure rotational (\ref{rot}) with a small tilt of the rotational plane.
But the motions (\ref{zwave}) with excited higher harmonics $n=2,3,\dots$
are non-trivial.

The transcendental equation (\ref{zfreq}) was studied in Ref.~\cite{PeSh}
where we proved that its roots $\om_n$ (with $\om_0=0$) and the functions
in Eqs.~(\ref{zwave})
$$
y_n(\s)= \cos(\om_n\s+\phi_n),\qquad\phi_n=\arctan(\om_n/Q_1),
\quad n=0,1,2,\dots
$$
are correspondingly the eigen-values and eigen-functions
of the boundary-value problem
\be
y''(\s)+\om^2y=0,\;\quad \om^2y(0)+Q_1y'(0)=0,\quad\om^2y(\pi)-
Q_2y'(\pi)=0.
\label{ybvp}\ee

It was proved in Ref.~\cite{PeSh} that the functions $y_n(\s)$,
$ n=0,1,2,\dots$ form the complete system in the class $C([0,\pi])$, and
the system $\exp(-i\om_n\tau)$, $n\in Z$ (with $\om_{-n}=-\om_n$) is complete
in the class of function $C(I)$, where $I=[\tau_0,\tau_0+2\pi]$.
So any continuous function $f(\tau)$ given in the segment $I$ may
be expanded in the Fouirer series
\be
f(\tau)=\sum_{n=-\infty}^{+\infty}f_n\exp(-i\om_n\tau),\qquad
\tau\in I=[\tau_0,\tau_0+2\pi].
\label{uFour}\ee

As was mentioned above, all information about any motion of this system
is contained in the function $U_i^\mu(\tau)$ given in the segment $I$.
If we expand any small disturbance $u_{1z}(\tau)$ or $u_{2z}(\tau)$ in the
segment $I$ into the Fourier series (\ref{uFour}), this series will describe
the evolution of the given disturbance for all $\tau\in R$, because
any term in (\ref{uFour}) satisfies the evolution equations (\ref{sysu}).
So any small disturbance of the rotational motion (\ref{rot}) in
$e_3^\mu$ direction may be expanded into the Fourier series with the
oscillatory modes (\ref{zwave}).

This also concerns the quasirotational motions in the rotational plane
$e_1,e_2$. They are determined by the first 4 equations (\ref{sysu}).
If we substitute $u_{i0}=A_i\exp(-i\tilde\om\tau)$,
$u_{ie}=B_i\exp(-i\tilde\om\tau)$ into this system we obtain the following
condition for existance of its non-trivial solutions:
\be
\!\left|\begin{array}{cccc}
Q_1-i\tilde\om &-Q_1 &Q_1+i\tilde\om q_{12} &-
Q_1\\ Q_1/v_1^2  &Q_1-i\tilde\om &-\tilde Q_2-
i\tilde\om q_{12} &Q_1-i\tilde\om \\
Q_1+i\tilde\om &Q_1 &\!(Q_1-i\tilde\om q_{12})\,e^{-2\pi
i\tilde\om}\!\! &Q_1 e^{-2\pi i\tilde\om} \\
\tilde Q_1+i\tilde\om q_{12}^{-1} &Q_2-i\tilde\om &
-Q_2v_2^{-2}e^{-2\pi i\tilde\om}&
(Q_2-i\tilde\om)\, e^{-2\pi i\tilde\om}
\end{array}\right|=0.
\label{deter}\ee
Here $q_{12}=Q_1/Q_2$, $\tilde Q_j=Q_j\kappa_j+Q_{3-j}+i\tilde\om$.

There are eigen-frequencies $\tilde\om=\tilde\om_n=n$, $n\in Z$ satisfying
this equation. But the correspondent functions
$u_i^\mu(\tau)=B_i^\mu\exp(-in\tau)$ after substitution into
Eqs.~(\ref{U+u}), (\ref{psdet}), (\ref{gen}) result in quasirotational
excitations of the motion (\ref{rot}), which may be obtained from
Eq.~(\ref{rot}) through the following reparametrization \cite{PeSh}
$$
\tilde\tau\pm\tilde\s=f(\tau\pm\tilde\s):\quad
f(\xi+2\pi)=f(\xi)+2\pi,\quad f'(\xi)>0,\quad\xi\in R
$$
on the same world surface. Eqs.~(\ref{eq})\,--\,(\ref{ort}) and the conditions
$\s_1=0$, $\s_2=\pi$ are invariant with respect to this reparametrization
so the oscillations with $u_i^\mu(\tau)=B_i^\mu\exp(-in\tau)$ have no
physical manifestations. They may be interpreted as ``longitudinal
oscillations" inside the string or, oscillations of the grid chart on the
world surface.

If we exclude these non-physical roots $\tilde\om_n=n$, the equation
(\ref{deter}) will reduce to following one:
\be
\frac{(\tilde\om^2-Q_1^2\kappa_1)(\tilde\om^2-Q_2^2\kappa_2)
-4Q_1Q_2\tilde\om^2}
{2\tilde\om\big[Q_1(\tilde\om^2-Q_2^2\kappa_2)+
Q_2(\tilde\om^2-Q_1^2\kappa_1)\big]}=\cot\pi\tilde\om,
\label{pfreq}\ee
One can numerate the roots $\tilde\om=\tilde\om_n$ of Eq.~(\ref{pfreq})
in order of increasing so that $\tilde\om_0=0$,
$n-1<\tilde\om_n<n$ for $n\ge1$ (and $\tilde\om_{-n}=-\tilde\om_n$).
The value $\tilde\om=\theta$ is also the root of Eq.~(\ref{pfreq}) but
it corresponds to the trivial quasirotational states, connected with
a small shift of the rotational center with respect to the coordinate origin.

In the following table some first roots $\om_n$ of Eq.~(\ref{zfreq})
and $\tilde\om_n$ of Eq.~(\ref{pfreq}) are represenred for the example of
the rotational motion (\ref{rot}) with $m_1=m_2$, $v_1=v_2=1/\sqrt2$,
$\th=0.5=Q_1=Q_2$:

\smallskip
\begin{center}
% \begin{table} \caption{ }
\begin{tabular}{|c|c|c|c|c|c|}\hline
$n$ & 1 & 2 & 3 & 4 & 5 \\ \hline
$\om_n$ & 0.5 & 1.2434 & 2.1457 & 3.102 & 4.078 \\ \hline
$\tilde\om_n$ & 0.9262 & 1.5 & 2.299 & 3.206 & 4.157 \\ \hline
\end{tabular}
%\label{table1}\end{table}
\end{center}

\smallskip

The roots $\tilde\om_n$ of Eq.~(\ref{pfreq}) are eigen-values
of the boundary-value problem
$$
y''(\s)+\om^2y=0,\;\quad (\om^2-Q_1^2\kappa_1)\,y(0)+2Q_1y'(0)=0,\quad
(\om^2-Q_2^2\kappa_2)\,y(\pi)-2Q_2y'(\pi)=0,
$$
similar to the problem (\ref{ybvp}). So the corresponding eigen-functions
and the system $\exp(-i\tilde\om_n\tau)$, $n\in Z$ is complete in the class
of function $C(I)$, $I=[\tau_0,\tau_0+2\pi]$ \cite{PeSh}.

Hence, an arbitrary quasirotational disturbance $u^\mu(\tau)$
may be expanded in the Fourier series similar to Eq.~{\ref{uFour}}.
Using this expansion for the disturbance (\ref{U+u}) $u_i^\mu$ of the
velocity vectors (\ref{Urot}) we obtain with the help of
Eqs.~(\ref{gen}), (\ref{psdet}) the following expression for an arbitrary
quasirotational motion of the string with massive ends \cite{Exc}:
\begin{eqnarray}
&\displaystyle
X^\mu(\tau,\s)=X^\mu_{rot}(\tau,\s)+\!\sum_{n=-\infty}^\infty
\Big\{e_3^\mu\alpha_n\cos(\om_n\s+\phi_n)\exp(-i\om_n\tau)&
\nonumber\\
&\qquad\qquad{}+\beta_n\big[e_0^\mu f_0(\s)+\acute e^\mu(\tau) f_\perp(\s)+
ie^\mu(\tau) \fpar (\s)\big]\exp(-i\tilde\om_n\tau)\Big\}.&
\label{osc}
\end{eqnarray}
Here the first term $X^\mu_{rot}$ describes the pure rotation (\ref{rot})
and
$$
\begin{array}{c}
f_0(\s)=\frac12(Q_1\kappa_1\tilde\om_n^{-1}-Q_1^{-1}\tilde\om_n)
\cos\tilde\om_n\s-\sin\tilde\om_n\s,\\
f_\perp(\s)=\Gamma_1(\Theta_n\tilde\om_n-h_nv_1)\,C_\theta C_\om
-v_1^{-1}C_\theta S_\om+\Gamma_1\theta\Theta_n S_\theta S_\om+h_n S_\theta
C_\om,\rule[3.3mm]{0mm}{1mm}\\
 \fpar (\s)=\Gamma_1(\Theta_n\tilde\om_n-h_nv_1)\,S_\theta S_\om
+v_1^{-1}S_\theta C_\om+\Gamma_1\theta\Theta_n C_\theta C_\om-h_n
C_\theta S_\om;\rule[3.3mm]{0mm}{1mm}
\end{array}
$$
$\kappa_i=1+v_i^{-2}$, $\displaystyle \;\Theta_n=\frac{2\theta}{\tilde\om_n^2-
\theta^2}$, $\displaystyle\;
h_n=\frac12\Big[\frac\theta{\tilde\om_n}\Big(\frac1{v_1}+v_1\Big)+
\frac{\tilde\om_n}\theta\Big(\frac1{v_1}-v_1\Big)\Big]$,
$\;C_\theta(\s)=\cos\theta\s$, $S_\theta(\s)=\sin\theta\s$,
$C_\om(\s)=\cos\tilde\om_n\s$, $S_\om(\s)=\sin\tilde\om_n\s$,
$\al_{-n}=\al_n^*$, $\beta_{-n}=-\beta_n^*$.

The quasirotational disturbances (\ref{osc}) with $\beta_n\ne0$ and $\al_n=0$
are small (if $\beta_n\ll\Omega^{-1}$) harmonic planar oscillations
or stationary waves in the rotational plane.
The shape of these stationary waves in the co-rotating frame of reference
(where the axes $x$ and $y$ are directed along $e^\mu$ and $\acute e^\mu$)
is approximately described by the function
$\beta_n\big[f_\perp(\s)-f_0(\s)\cos(\theta\s+\phi_1)\big]$
if the deflection is maximal.
For each $n$ this rotating curved string oscillates at the frequency
$\Omega_n=\Omega\tilde\om_n/\theta$, it has $n$ nodes in $(0,\pi)$
(which are not strictly fixed because $f_0$ and $\fpar $ are non-zero) and the
moving quarks are not in nodes. Note that Eq.~(\ref{osc}) describes
both the deflection of the endpoints
$\beta_n\acute e^\mu f_\perp(\s_i)\sin\om_n\tau$ and
their radial motion $\beta_ne^\mu \fpar (\s_i)\cos\om_n\tau$.

The frequencies $\om_n$ and $\tilde\om_n$ from Eqs.~(\ref{zfreq}) and
(\ref{pfreq}) are real numbers,
so the rotations (\ref{rot}) of the string with massive ends are stable
in the linear approximation.
Hence, one may consider the Fourier series (\ref{osc})
for an arbitrary quasirotational motion as the basis for quantization
of some class of motions (quasiro\-tational states) of the string with massive
ends in the linear vicinity of the solution (\ref{rot}).
Note that for the string baryon models ``three-string" (Y) and the linear
configuration ($q$-$q$-$q$) their rotational motions are unstable,
the analogs of Eq.~(\ref{pfreq}) for these models contain complex frequencies
\cite{Exc,Exclin}.

The quasirotational states (\ref{osc}) of the relativistic string with massive
ends may be used for describing radial excitations of hadrons. In this connection
the most interesting states (\ref{osc}) are oscillations in the rotational
plane (the planar modes) with $\al_n=0$, $\beta_n\ne0$, and among them ---
the main planar mode with $n=1$. Let us consider them in detail and calculate
the energy $E$ and angular momentum $J$ for these excitations.

\bigskip
\noindent{\bf 3. Energy $E$ and angular momentum $J$ of quasirotational states}
\medskip

The main planar quasirotational mode (\ref{osc}) with $n=1$ is
non-trivial\footnote{Unlike the $n=1$ mode
of orthogonal oscillations (\ref{zwave}).}. If only this mode is excited  the
motion is quasiperidical, the string at the frequency $\Omega_1=
\Omega\tilde\om_1/\theta$ ($1.5\Omega<\Omega_1<2\Omega$) slightly deflects
from pure uniform rotation keeping almost rectilinear shape. The length of the
string (distance between quarks) varies in accordance with this deflection
at the frequency $\Omega_1$. The endpoints draw curves close to ellipses both
in the co-rotating frame of reference (the pure uniform rotational position
of an end is in the center of this ellipse) and in the frame of reference $Oxy$.
In the latter case this ellipse (close to a circle) rotates in the main
rotational direction because the frequencies
$\Omega_1$ and $2\Omega$ are incommensurable numbers: $\Omega_1<2\Omega$.
The more this disturbance (its amplitude $\beta_1$) the more those rotating
ellipses differ from circles.

For the second oscillatory mode ($n=2$) the middle part of the rotating string
oscillates at the frequency $\Omega_2$, but deviations of the endpoints are
small in comparison with this amplitude in the middle part. This motion is shown
in details in Ref.~\cite{stabPRD}.

Possible applications of these string states in hadron spectroscopy depend on
their physical characteristics. Let us calculate the most important among them:
the energy $E$ and angular momentum $J$ of the quasirotational state (\ref{osc}). For an arbitrary classic state of the relativistic string
with massive ends they are determined by the following integrals
(Noether currents) \cite{BN}:
\bea
&\displaystyle
P^\mu=\int\limits_{\s_1}^{\s_2} p^\mu(\tau,\s)\,d\s
+p_1^\mu(\tau)+p_2^\mu(\tau),\qquad
p^\mu(\tau,\s)=\gamma\frac{(\dot X,X') X^{\pr\mu}-X'{}^2\dot X^\mu}
{\big[(\dot X,X')^2-\dot X^2X^{\pr2}\big]^{1/2}},& \label{Pimp}\\
&\displaystyle
{\cal J}^{\mu\nu}={\!\!\int\limits_{\s_1}^{\s_2}\!}\Big[X^\mu(\tau,\s)\,
p^\nu(\tau,\s)-X^\nu(\tau,\s)\,p^\mu(\tau,\s)\Big]\,d\s+\sum_{i=1}^2
\Big[x_i^\mu(\tau)\,p_i^\nu(\tau)-x_i^\nu(\tau)\,p_i^\mu(\tau)\Big],&
\label{Mom}\eea
where $p_i^\mu(\tau)=m_iU_i^\mu(\tau)$, $s_1=0$, $s_2=\pi$.
In the orthonormal gauge (\ref{ort})
$p^\mu(\tau,\s)=\gamma\dot X^\mu(\tau,\s)$.

The square of energy $E^2$ equals the scalar square of the conserved
$R^{1,3}$-\,vector of momentum (\ref{Pimp}): $P^2=P_\mu P^\mu=E^2$.
In the center of mass reference frame $E=P^0$, the latter case takes place
for the expression (\ref{osc}).

If we substitute the Fourier series (\ref{osc}) into expression (\ref{Pimp})
we'll obtain the following equality for the 4-momentum:
\be
P^\mu=P^\mu_{rot}=e_0^\mu\sum_{i=1}^2[\gamma\Omega^{-1}\arcsin v_i+
m_i\Gamma_i].
\label{Prot}\ee

In other words, the energy of an arbitrary quasirotational state (\ref{osc})
equals the energy $E_{rot}$ (\ref{EJ}) of the pure rotational motion
(\ref{rot}), all quasirotational modes with amplitudes $\al_n$, $\beta_n$ yield
zero contributions.
This result is obtained in the linear approximation with respect to
$\al_n$, $\beta_n$ in the expressions for the endpoints' momenta
$p_i^\mu=m_iU_i^\mu$, for example
$$
U_1^\mu(\tau)\simeq U^\mu_{1(rot)}+\Omega v_1\Gamma_1^2\sum_{n\ne0}\Big\{
\frac{e_3^\mu\al_n}{\mbox{\footnotesize$\sqrt{\tilde\om_n^2+Q_1^2}$}}
e^{-i\om_n\tau}
-i\beta_n\big[e_0^\mu +v_1^{-1}\acute e^\mu(\tau) +
ih_ne^\mu(\tau)\big]\,e^{-i\tilde\om_n\tau}\Big\}.$$
Here $U^\mu_{i(rot)}$ is the expression (\ref{Urot}).
In this approximation the contribution of the ends in Eq.~(\ref{Pimp})
exactly compensates the string contribution
$\int p^\mu(\tau,\s)\,d\s$ for each oscillatory mode.

This property is similar to vanishing contributions of high oscillatory modes
in the Fourier series for the massless open string \cite{BN}. In the latter case
energy of these oscillation may be found from the orthonormality conditions (\ref{ort}) (the Virasoro conditions).

In the case of quasirotational states (\ref{osc}) we have no the Virasoro
conditions, because the orthonormality conditions (\ref{ort}) were previously
solved and taken into account in expressions (\ref{rot}), (\ref{psdet}),
(\ref{sysU}), (\ref{U+u}) in the linear approximation with respect to $u_i^\mu$.

The equality (\ref{Prot}) of the energies of the quasirotational $E$ and
pure rotational motion $E_{rot}$ (compare with the similar result in
Ref.~\cite{AllenOV}) looks questionable: we always can add a disturbance with
nonzero energy $\Delta E$ to the pure rotation (\ref{rot})
$X^\mu_{rot}(\tau,\s)$. But there is no contradiction: the resulting motion
will be the quasirotational state (\ref{osc}) with respect to the rotation
(\ref{rot}) with the energy $E_{rot}+\Delta E$.

The classical angular momentum of the quasirotational motion (\ref{osc})
is determined by Eq.~(\ref{Mom}) and in the case\footnote{Remind that inequality
$\al_1\ne0$ results in a trivial tilt of the rotational plane.} $\al_1=0$
it takes the form
\be
{\cal J}^{\mu\nu}=j_3^{\mu\nu}\bigg\{J_{rot}-\gamma\!\sum_{n=1}^\infty\!
|\beta_n|^2\frac{\Gamma_1^2\th^2Z_{1,n}}{\tilde\om_n^2(\tilde\om_n^2-\th^2)}
\bigg[\pi\th(3\tilde\om_n^2+\th^2)+\sum_{i=1}^2\frac1{v_i\Gamma_i}
\Big(\frac{4\tilde\om_n^2}{Z_{i,n}}-v_i^2(\tilde\om_n^2-\th^2)\Big)
\bigg]\bigg\}
\label{Jqrot}\ee
Here $j_3^{\mu\nu}=e_1^\mu e_2^\nu-e_1^\nu e_2^\mu=e^\mu\acute e^\nu-
e^\nu\acute e^\mu$, $\displaystyle Z_{i,n}=1+\frac{(\tilde\om_n^2-\th^2)^2}
{4v_i^2\Gamma_i^2\th^4}$.

Note that the 2-nd order contribution $\Delta J$ (proportional to $|\beta_n|^2$)
in Eq.~(\ref{Jqrot}) to the momentum $J_{rot}$ (\ref{EJ}) of the pure
rotational motion is always negative. This is natural, the rotational motion
(\ref{rot}) has the maximal angular momentum among the motions with given energy
\cite{BN}. And we consider the quasirotational states (\ref{osc}) with fixed
energy (\ref{Prot}) $E=E_{rot}$.

This result shows the possibility to apply these states (especially the state
with $\beta_1\ne0$) for describing high radial excitations of mesons and
baryons. If for a planar excited mode the correction $\Delta J=-1$
(in the natural units $\hbar$) we may interpret this state as the radial
excitation lying on the first daughter Regge trajectory. It has the same
energy $E_{rot}$, so it may be only the excitation of the state with lower
energy on this daughter trajectory.

Hence in this approach the slope of radial trajectories is equal to that
for the orbital Regge trajectories because the increase of energy is the same
for enlarging the momentum $J$ or the radial quantum number $n$.
After developing the quantization procedure on the basis of the Fourier series
(\ref{osc}) this method may be applied for describing higher radial
excitations in hadron spectroscopy.

\bigskip
\noindent{\bf Conclusion}
\medskip

For the obtained class of quasirotational motions for the relativistic string
with massive ends the possible applications are investigated.
It is shown that the quasirotational states (\ref{osc}) may be used for
describing radial excitations of mesons and baryons (in the frameworks of the
model $q$-$qq$). This possibility may be brought about after developing
the quantization procedure for the quasirotational states (\ref{osc}).
This procedure is the object for further study. It will include the usual
interpretation of the amplitudes $\al_n$, $\beta_n$ in this Fourier series
as quantum operators with defining their commutability relations.

Note that the expressions (\ref{Prot}) and (\ref{Jqrot}) for the energy and
angular momentum are obtained in the linear approximation with respect to
small disturbances $u_i^\mu(\tau)$ in Eq.~(\ref{U+u}). If these disturbances
are not small the second order corrections may change Eqs.~(\ref{Prot}),
(\ref{Jqrot}). But these second order corrections can not be determined
uniquely from the given linear disturbances $u_i^\mu(\tau)$ or
(this is equivalent) $\al_n$, $\beta_n$. These corrections
$\stackrel{2}{u}{\!}_i^\mu(\tau)$ have to satisfy only one condition
--- the equality $U_i^2(\tau)=1$, where
$U_i^\mu(\tau)= U_{i(rot)}^\mu(\tau)+u_i^\mu(\tau)+
\stackrel{2}{u}{\!}_i^\mu(\tau)$, and other degrees of freedom in
a choice of $\stackrel{2}{u}{\!}_i^\mu(\tau)$ are not fixed. So one can choose
the correction  $\stackrel{2}{u}{\!}_i^\mu$ resulting in the same value of
energy (\ref{Prot}).

This approach was developed for the string with massive ends.
We are to emphasize that it is not applicable for more complicated
string baryon models $q$-$q$-$q$ and Y (``three-string") because
the rotational motions for them are unstable and we can't quantize states
in their linear vicinity \cite{stabPRD,Exc,Exclin}.

\smallskip

The work is supported by the Russian Foundation of Basic Research,
grant  00-02-17359.

\end{document}